# DETECTING SCHIZOPHRENIA WITH 3D STRUCTURAL BRAIN MRI USING DEEP LEARNING

**Junhao Zhang**[1]   **Vishwanatha M. Rao**[1]   **Ye Tian**[1]   **Yanting Yang**[1]   **Nicolas Acosta**[1]   **Zihan Wan**[2]
**Pin-Yu Lee**[1]   **Chloe Zhang**[8]   **Lawrence S. Kegeles**[3,5]   **Scott A. Small**[6,7]   **Jia Guo**[3,4*]

[1]Department of Biomedical Engineering, Columbia University, New York, NY, USA
[2]Department of Applied Mathematics, Columbia University, New York, NY, USA
[3]Department of Psychiatry, Columbia University, New York, NY, USA
[4]The Mortimer B. Zuckerman Mind Brain Behavior Institute, Columbia University, New York, NY, USA
[5]Department of Radiology, Columbia University, New York, NY, USA
[6]Department of Neurology, Radiology, and Psychiatry, Columbia University, New York, NY, USA
[7]The Taub Institute for Research on Alzheimer's Disease and the Aging Brain, Columbia University, New York, NY, USA
[8]Jericho High School, Jericho, NY, USA
[*]Correspondence: jg3400@columbia.edu

June 24, 2022

## ABSTRACT

Schizophrenia is a chronic neuropsychiatric disorder that causes distinct structural alterations within the brain. We hypothesize that deep learning applied to a structural neuroimaging dataset could detect disease-related alteration and improve classification and diagnostic accuracy. We tested this hypothesis using a single, widely available, and conventional T1-weighted MRI scan, from which we extracted the 3D whole-brain structure using standard post-processing methods. A deep learning model was then developed, optimized, and evaluated on three open datasets with T1-weighted MRI scans of patients with schizophrenia. Our proposed model outperformed the benchmark model, which was also trained with structural MR images using a 3D CNN architecture. Our model is capable of almost perfectly (area under the ROC curve = 0.987) distinguishing schizophrenia patients from healthy controls on unseen structural MRI scans. Regional analysis localized subcortical regions and ventricles as the most predictive brain regions. Subcortical structures serve a pivotal role in cognitive, affective, and social functions in humans, and structural abnormalities of these regions have been associated with schizophrenia. Our finding corroborates that schizophrenia is associated with widespread alterations in subcortical brain structure and the subcortical structural information provides prominent features in diagnostic classification. Together, these results further demonstrate the potential of deep learning to improve schizophrenia diagnosis and identify its structural neuroimaging signatures from a single, standard T1-weighted brain MRI.

## 1   Introduction

Schizophrenia is a progressive neuropsychiatric disorder that is characterized by structural changes within the brain. Recent findings from a large meta-analysis suggest that schizophrenia is associated with gray matter reductions across multiple subcortical regions including the hippocampus, amygdala, caudate, and thalamus, with structural changes in shape within those regions supporting changes in functional brain networks [1]. In addition to the altered shape of such brain structures, schizophrenia is also associated with significantly greater mean volume variability of the temporal cortex, thalamus, putamen, and third ventricle [2]. Other studies also affirm the enlargement of ventricles in schizophrenia [3, 4]. While gray matter reductions are most consistently reported in the subcortical regions, reductions have also been identified in areas such as the prefrontal, temporal, cingulate, and cerebellar cortices [5, 6]. Loss of gray matter volume has been shown to not only mark the onset of schizophrenia but also progress alongside the illness [7].

Despite these documented changes, accurate and rapid detection of schizophrenia remains a pressing problem; previous studies are limited to only characterizing structural abnormalities at a group level, with no concrete method to make individual diagnoses at a subject level. Additionally, the diagnosis of schizophrenia based on DSM-5 criteria is costly both in terms of time and resources, without ensuring objectivity. Therefore, it is imperative to develop an objective screening tool to diagnose schizophrenia and potentially improve patient prognosis by allowing for earlier intervention.

Various attempts have been proposed to take advantage of the structural alterations present in schizophrenia for classification using neuroimaging data. Machine learning algorithms have historically presented the ability to classify psychiatric disorders in this manner [8, 9]. In particular, the support vector machine (SVM), a supervised learning algorithm able to capture non-linear patterns in high-dimensional data, has been most prevalent in schizophrenia classification. Other popular machine learning algorithms for schizophrenia classification include multivariate pattern analysis, linear discriminant analysis, and random forest [8, 10]. While standard machine learning approaches have demonstrated compelling results, their performance highly depends on the validity of manually extracted features [8]. Such features are traditionally extracted based on a combination of previously known disease characteristics and automatic feature selection algorithms [11]. These features may not completely encode the subtle neurological differences associated with schizophrenia; alternatively, they may encode too much unnecessary information requiring additional feature reduction [12].

Deep learning has recently emerged as a new approach demonstrating superior performance over standard machine learning algorithms to classify neurological diseases using structural MRI data. Specifically, Convolutional Neural Networks (CNNs) can learn and encode the significant features necessary for classification and have become popular in medical image analysis [13-15]. This property makes CNNs uniquely suited to tasks like schizophrenia classification, where the specific features selected can dramatically impact model performance. Some studies have already demonstrated the utility of CNNs for schizophrenia classification. Oh J *et al.*, [16] achieved an impressive state-of-the-art performance (area under the ROC curve = 0.96) using 3D CNN for schizophrenia classification based on structural MRI data and was thus compared to as the benchmark model. Nevertheless, they struggled to generalize well on an unseen private dataset. Their inconsistent performance may be attributed to the dataset and patient variability as well as certain pre-processing choices, such as the inclusion of whole head as opposed to whole-brain MRI data and severe downsampling. Moreover, their region of interest analysis was limited and did not investigate brain structures in depth to inform specific changes in structural features associated with schizophrenia. Hu et al. combined structural and diffusion MRI scans for schizophrenia classification and found that 3D CNN models could outperform 2D pre-trained CNN models as well as multiple standard machine learning algorithms like SVM. Despite this, their best 3D model only reached the area under the ROC curve of 0.84 [17]. As a consequence, though deep learning has advanced neuroimaging-based schizophrenia classification, the preprocessing and acquisition of large datasets coupled with the achievement of high model performance and generalization remains a great challenge.

In this study, we not only address the limitations in schizophrenia classification with T1-weighted (T1W) MRI data but also take advantage of class activation maps (CAM) in a deep learning network to visualize informative regions with disease vulnerability. Our main contributions include the following: firstly, we develop a 3D CNN using structural MRI scans to yield a performance better than the benchmark model [16] for schizophrenia classification; and secondly, we apply gradient class activation maps to localize the brain regions related to schizophrenia identification. By visualizing feature activations, we provide further evidence that the structure of subcortical regions and ventricular areas [1,2] are affected in schizophrenia.

## 2 Methods

### 2.1 Study Design

For our experiments, firstly, we implemented the schizophrenia classification task with the benchmark model using 887 structural whole-head (WH) T1W scans, following the same pre-processing and parameter settings as the implementation in the CNN benchmark [16]. Secondly, a modified 3D VGG [18] with squeeze excitation (SE) [19] and batch normalization (BN) [20] model (SE-VGG-11BN) was used to perform the schizophrenia vs. cognitive normal binary classification task with the input of 887 T1W structural whole-brain (WB) scans. Of the 887 scans, 437 were controls and 450 were schizophrenia patients.

### 2.2 Data Selection

The neuroimaging data used in this study from patients with schizophrenia and normal subjects were downloaded from the SchizConnect database (http://schizconnect.org/). Data from three studies, COBRE [21], BrainGluSchi [22], and NMorphCH [23], were collected and organized in this public database. Images not applicable for training the deep network (e.g., those with excessive motion or noise or an image error) were excluded by visual inspection. In our experiment, the scans among all 3 studies were acquired from the same clinical MRI scanner model (SIEMENS Trio) using a standard 3D MPRAGE sequence with isotropic 1 mm resolution at 3T field strength. The data from these studies were high in quality and resolution, and the data acquisition time was relatively recent, varying from 2008 to 2013. In summary, the data in these studies were abundant and appropriate for model training. More detailed information about this data is illustrated in Fig. 1A.

**A**

| Scan Parameters | BrainGluSchi | COBRE | NMorphCH |
|---|---|---|---|
| Scanner | SIEMENS TrioTim | SIEMENS TrioTim | SIEMENS TrioTim |
| Field strength | 3T | 3T | 3T |
| Sequence | MPRAGE | MPRAGE | MPRAGE |
| Voxel size (mm) | 1.0x1.0x1.0 | 1.0x1.0x1.0 | 1.0x1.0x1.0 |
| TR/TE (msec) | 2530/1.64 | 2530/1.64 | 2400/3.16 |
| Acquisition year | 2010 - 2013 | 2009 - 2013 | 2008 - 2013 |
| Number of control scans (Train/Validation/Test) | 89 (73/8/8) | 237 (199/22/16) | 111 (89/10/12) |
| Number of patient scans (Train/Validation/Test) | 86 (70/7/9) | 243 (207/23/13) | 121 (98/6/17) |
| Female % | 20.3% | 24.0% | 40.0% |
| Age range (year) | 16-66 | 18-66 | 19-46 |
| Age mean ± SD | 37.6 ± 13.3 | 38.3 ± 12.6 | 32.2 ± 7.6 |

**B**

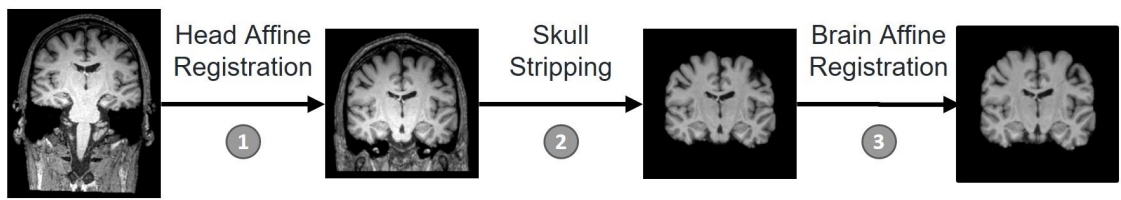

**Figure 1: Sample characteristics, distribution of public schizophrenia MRI datasets and the preprocessing pipeline.** A. Acquisition parameters of the T1W MRI scans and the patient demographic information of each dataset. In the BrainGluSchi, COBRE, and NMorphCH datasets, normal scans consisted of whole head structural T1W MR images obtained from healthy control subjects and schizophrenia scans consisted of whole head structural T1W MR images obtained from schizophrenia and schizoaffective disorder patients. B. Data preprocessing pipeline to generate the input. For each structural MRI, we process the T1W 3D volume through a standardized pipeline consisting of three steps: (1) whole head T1W affine registration to the MNI152 template space, (2) skull stripping, and (3) whole brain affine registration to the MNI152 template space. The preprocessing of structural T1W MR data is necessary to remove unwanted artifacts and transform the data into a standard format before training the deep learning models.

### 2.3 Data Pre-processing

Optimizing a deep learning model using data in this specific space requires the algorithm to learn discriminative patterns when the samples are in large numbers and include all of the expected variations. By pre-processing our images, we could alleviate some of the confounding factors, enabling the model to handle the entire image at once and automatically determine the most important task-related pattern in the data.

In our data preprocessing pipeline, firstly, the raw whole-head scans from three studies were registered to the MNI152 unbiased template by robust affine registration [24, 25], which is denoted by step one. Following the whole-head scans' registration, skull-stripping was applied on the whole-head scans using the Brain Extraction Tool [26] to obtain whole-brain (WB) MRI T1W scans, denoted by step two. After that, we affine-registered these whole-brain MRI T1W scans to the MNI152 unbiased template, denoted by step three. The details of these steps are illustrated in Fig. 1B.

Through affine registration, the MRI T1W scans kept similar structures in roughly the same spatial location using one template as the gold standard. We thereby reduced the variance in brain features, such as the brain volume, while still preserving differences in local anatomy, which may presumably reflect schizophrenia-related effects on brain structures. This operation could thus enable the model to focus on the decision-making patterns underlying the data.

After visual inspection of the preprocessed scans and removal of low-quality scans to avoid their potential negative effects on the classification task, the prepared data with 887 WB MRI T1W scans were selected and randomly assigned to 10 subsets. Each subset contained a similar number of samples. Randomization was performed on the subject level to prevent data leakage. To train and evaluate the model, eight out of ten subsets were randomly selected to make up the training set. Of the other two subsets, one was used as the validation set while the other was used as the test set. Consequently, the dataset was partitioned into the train/validation/test dataset by a ratio of approximately 8:1:1 at the subject level. The gender and age distribution in each subset were similar.

Down-sampling (×2) was applied to the input data with a matrix size of $192\times192\times192$ to help preserve the image information while extending the possible training batch size. This operation aimed to achieve a balance between resolution and batch size. For the model, the input is the processed 3D whole-brain T1W MRI scan while the output is a continuous-valued number representing the predicted schizophrenia likelihood.

### 2.4 Model Architecture and Implementation

For the schizophrenia classification tasks with one single input modality, the architecture 3D "VGG-11 with batch normalization" (Fig. 3) adapted from "VGG-19BN" [27] was developed in the PyTorch platform. VGG models are standard deep CNN architectures with 5 convolutional blocks proposed by the Visual Geometry Group, Oxford University [18]. They represent a group of robust architectures for object detection and image classification. This modified 3D VGG model with batch normalization and squeeze-and-excitation block (SE-VGG-11BN) was composed of two basic components: a feature extractor and a classifier. In the feature extraction portion, there was one down-sampling operation followed by five 3D convolution blocks, with each block containing 3D convolution, 3D batch normalization (BN), 3D squeeze-and-excitation (SE) operation, and 3D max-pooling.

Details of the operations involved in the convolutional block are illustrated as follows. The kernel size is $3\times3\times3$ and the padding and stride number is 1 in the 3D convolution. 3D batch normalization (BN) follows the convolution operation and normalizes inputs to layers in a neural network for each mini-batch. By rescaling and recentering, BN reduces the internal covariate shifts, enabling a higher learning rate. One notable difference between our model and the common VGG model lies in the introduction of the squeeze-and-excitation operation, which is a channel-wised attention mechanism used to improve the representational power of a CNN network. It adds weights factors to channels and accordingly recalibrates them to enhance significant features while ignoring the irrelevant features at almost no additional computational cost in the existing architecture. The channel-to-channel ratio is the only hyperparameter, which was tuned in the range from 8 to 32 (including 8, 12, 16, …, and up to 32), and set at 16 in the 3D SE operation. In the max-pooling, the kernel size and stride are $2\times2$ and 2, respectively. One slight difference from previous 3D convolution blocks is that the max-pooling in the last convolution block is abandoned since we needed a larger receptive field to generate the class activation map. In the classifier portion, three dense and two dropout layers are used to constitute the linear mapping. All the activation functions in feature extraction and classifier are rectified linear units (ReLU) [28] except for the penultimate and final activations, which are sigmoid and softmax functions respectively. The details of the proposed model are illustrated in Fig. 6.

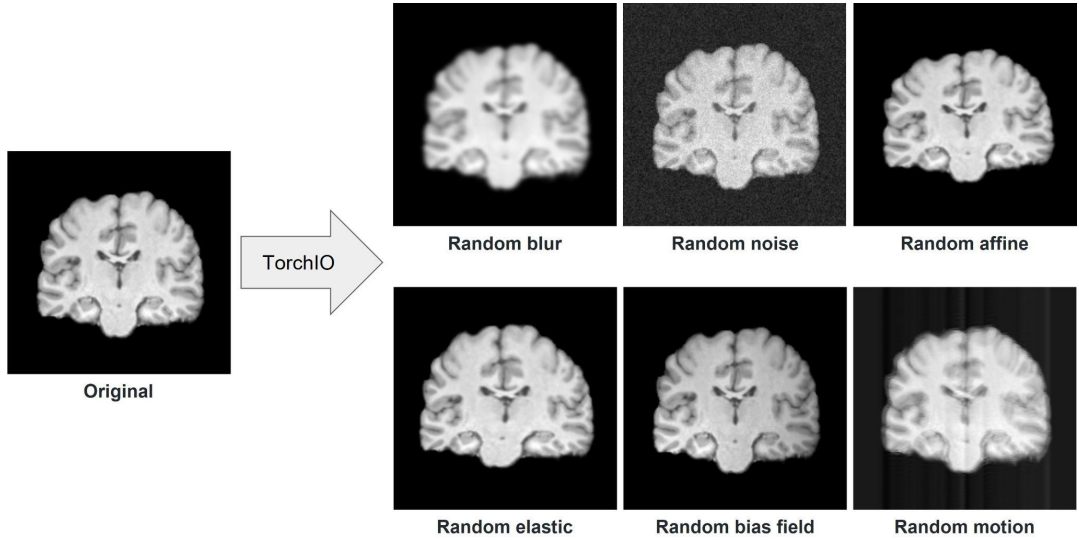

**Figure 2: An example of 3D T1-weighted MRI data augmentation results in the coronal view.** From left to right: the original single coronal slice; the single slice after randomly blurring; after adding Gaussian noise; after applying random affine transformation and resampling; after applying random dense and elastic deformation; after applying random bias field distortion; and after adding random motion artifact, respectively.

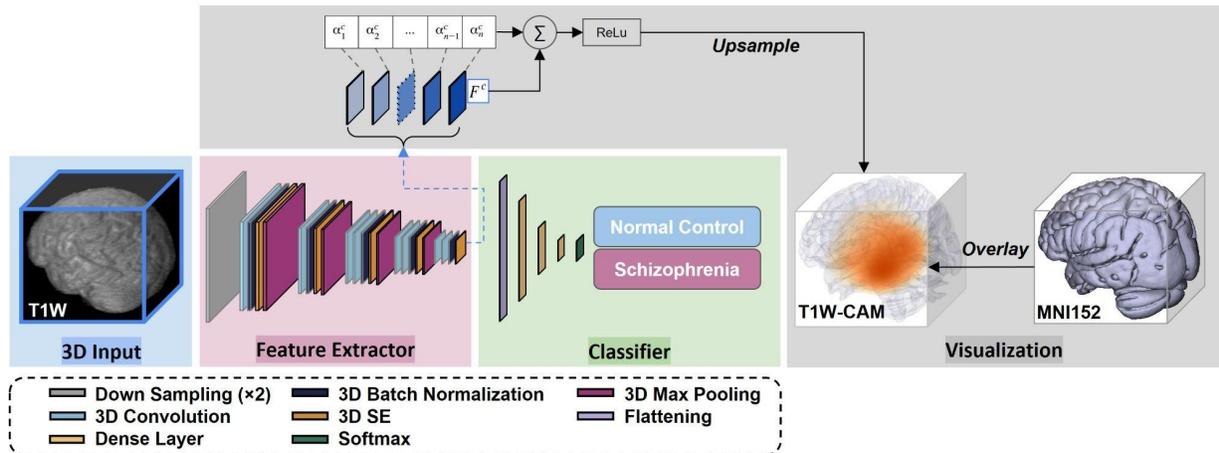

**Figure 3: The flow of the classification and visualization in the 3D SE-VGG-11BN CNN model.** This model consists of a modified 3D VGG-11 network with squeeze-and-excitation (SE) block and batch-normalization (BN) using T1W MRI as the model input. The class of one given T1W scan is predicted by two steps in the model: 1) extracting hierarchical features, and 2) classifying these features. In the feature extractor portion, the data is firstly under-sampled x2 and goes through several convolution blocks consisting of 3D convolution, 3D batch normalization, 3D max pooling, and 3D SE operation. The classifier consisting of three dense layers with dropout regularization yields the final prediction result. The classifier consisting of three dense layers with dropout regularization yields the final prediction result. In the feature extractor part, feature maps generated by filters at the last convolution layer are shown. These feature maps are used for visualization through the generation of the class activation map by weighting them with channel-wise average gradients.

In the training phase, the initial learning rate was set to 1e-4 (tuned in the range from 1e−3 to 1e−6) and the batch size was 5. The setting of batch size was chosen considering convergence speed and the memory limit. The loss function was the cross-entropy loss, and the Adam [29] method was used to optimize the model parameters. An early stopping strategy was introduced to the training phase to avoid over-fitting. The number of epochs was set to 300.

Data augmentation could help improve the model performance by making the model more agnostic to subject-level variation. As a result, a data augmentation strategy was used in our model training as well. Basic transformers from the TorchIO library were imported to transform the raw data and thus increase the number of training datasets. The data in the training set would go through random blurring with a probability of 0.1 and random noise addition with a probability of 0.6. After these two transformations, the data would go through one of the following with a probability of 0.2: random affine transformation with scaling factor = 1 or random elastic deformation. Finally, the image would undergo random bias field distortion with a probability of 0.1 followed by random motion distortion with a probability of 0.05. Examples of each of these transformations applied on T1W WB MRI scans are shown in Fig. 2.

### 2.5 Performance Evaluation of the Model

To evaluate the descriptiveness of the predicted schizophrenia likelihood, we conducted receiver-operating characteristics (ROC) studies to analyze the concordance between the model-generated classification and the ground truth labels. The ROC curves, one for each trained classification model, represent the classification performances at each potential numerical threshold to binarize the predicted schizophrenia likelihood score. The sensitivity and specificity (the sum of which peaks at the operating point), as well as the area under the ROC curve (AUC), demonstrate the effectiveness of the classification method. The significance of the difference among these ROC curves was calculated using DeLong's test [30].

### 2.6 Generalization Evaluation of the Model

To demonstrate the generalization of the models, data from COBRE and NMorphCH studies were selected to train the model, and data from BrainGluSchi with a nearly similar acquisition configuration was used for evaluating the capability of model generalization. The same training strategies and hyperparameter settings were maintained in the experiment.

### 2.7 Explainability of the Model with Grad-CAM

To validate the models, a gradient class activation map (Grad-CAM) was introduced to our experiment to check whether the model focuses on task-related patterns instead of some irrelevant information in the data. After excluding the possibility of the model focusing on meaningless regions in the data by applying a rough brain mask, we further investigated the brain regions that had the most contributions to the schizophrenia classification task by visualizing the class activation maps (CAM) [31]. By visualizing feature activations, we could identify which regions of the input images contribute to the classification results. We used the WB T1W scans from all the subjects with schizophrenia to generate an averaged CAM for the schizophrenia class. We had a great interest in whether the brain regions the classifier found the most relevant to the schizophrenia class were physiologically meaningful. The weighted feature maps only activate features that have a positive influence on the prediction after applying ReLU nonlinearity operations. The heatmap of the last convolution layer highlights the most important region for classifying the sample, whereas the maps for shallow layers localize more fine-grained features.

## 3 Results

While training, we found that SE-VGG-11BN converged faster than the benchmark model on the training set and performed better than the benchmark model on the validation set. After training both our model and the benchmark model, we tested them on the same stand-alone set of scans, 51 with schizophrenia and 49 without schizophrenia. The SE-VGG-11BN model using structural T1 WB scans exhibited better performance than the benchmark model across all metrics (0.921 accuracy, 0.949 sensitivity, and 0.946 specificity). The quantitative performance metrics are summarized in Fig. 4B and C. When inspecting the ROC curves (Fig. 4A), we found that the SE-VGG-11BN model with the input of structural T1 WB scans achieved 0.987 AUC, which outperforms the benchmark model that achieved 0.938 AUC. The $p$-value of the ROC test (DeLong's test) indicated our model is significantly better than the benchmark model using structural T1 WH scans at a level of 0.05.

SE-VGG-11BN also demonstrated improved generalization performance. When the BrainGluSchi dataset was used only for testing and the COBRE and NMorphCH datasets were used for training and validation, we observed significantly superior AUC performance of SE-VGG-11BN (0.913) over the benchmark model (0.810) (Fig. 4D). These results validate the generality of our model and highlight its reliability in predicting schizophrenia on unseen and heterogeneous structural MRI data.

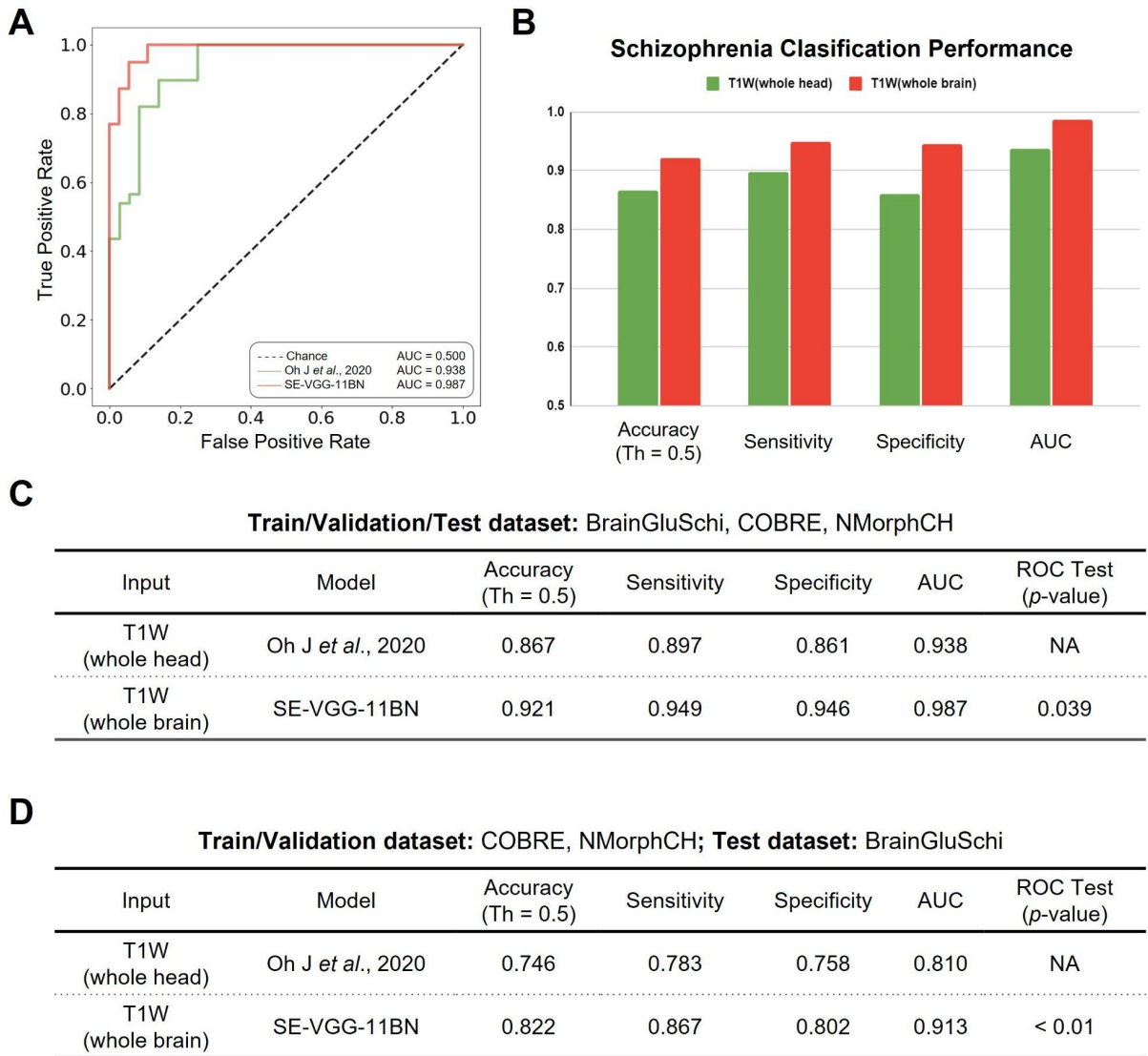

**Figure 4: Quantitative performance comparisons of our model and the benchmark model.** A. Receiver operating characteristics (ROC) curves for schizophrenia classification on the dataset. The green line represents the ROC curve of the benchmark model with the input of T1W WH scans. The red line represents the ROC curve of the SE-VGG-11BN with the input of T1W WB scans. B. Bar plot of the classification performance of these models in terms of the accuracy (at the default operating threshold of 0.5, Th = 0.5), sensitivity, specificity, and the area under the ROC curve (AUC). C. Table quantitatively summarizes the performance of these models. The $p$-value of the ROC test (DeLong's test) indicated our model is significantly better than the benchmark model at a level of 0.05. D. Generalizability of the two models trained by COBRE and NMorphCH datasets on unseen BrainGluSchi test dataset. Considering components A through D, SE-VGG-11BN exhibits a significantly better performance than the benchmark model.

To investigate the most pertinent spatial features contributing to the classification ability of the proposed deep learning algorithm, we further analyzed regional information in the anatomical structures of the structural MRI data. We illustrate the class activation map of the SE-VGG-11BNl for schizophrenia patients in Fig. 5, localizing discriminative regions for schizophrenia classification in the sagittal, axial and coronal views. The class activation map indicates large "activation" in the subcortical and ventricular regions, suggesting the importance of these regions in differentiating schizophrenia for our proposed model.

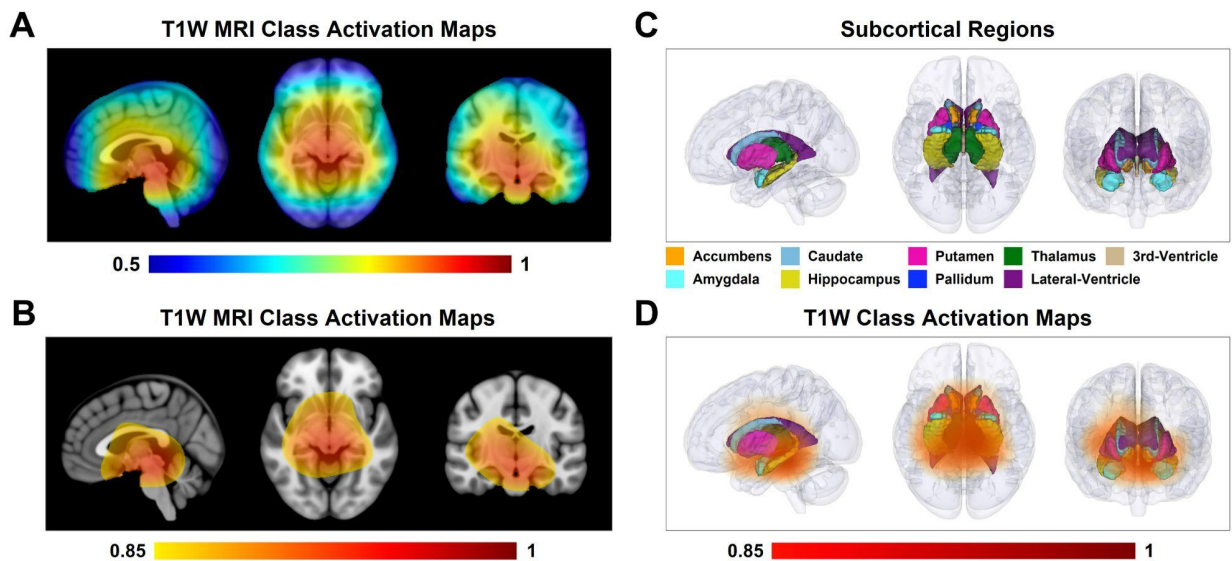

**Figure 5: Class activation map on T1-weighted brain MRI images in schizophrenia classification.** A. The class activation map (CAM) derived from feature maps in the last convolution layer from schizophrenia patients localizes the discriminative regions for schizophrenia classification in the sagittal, axial and coronal views. The color bar ranges from 0.5 to 1. The higher the value, the more important role the region plays in schizophrenia classification. B. The CAM is displayed in the sagittal, axial, and coronal views with a threshold of 0.85. The color bar ranges from 0.85 to 1. The thresholded CAM primarily lies in the subcortical regions and ventricular areas. C. Subcortical regions and ventricular areas are visualized in 3D in the sagittal, axial, and coronal views. D. The 3D volume rendering of the thresholded CAM demonstrates the location of the most discriminative regions in the sagittal, axial, and coronal views. The color bar ranges from 0.85 to 1. The thresholded CAM covers the subcortical regions and the ventricular areas.

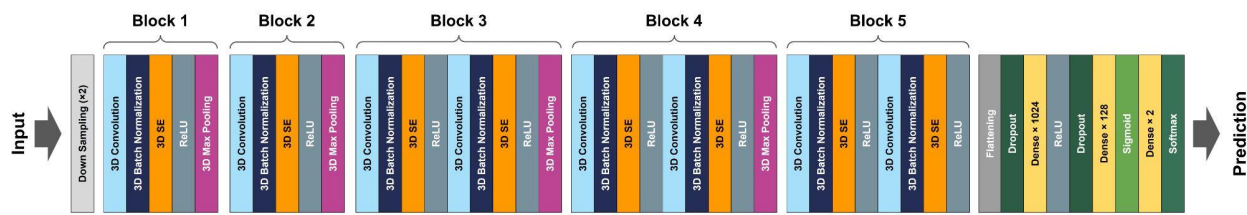

**Figure 6: Overall network architecture of the proposed approach.** The inputs are 3D brain volumes after being down-sampled by a factor of 2. This modified 3D VGG model is composed of two basic components: a feature extractor and a classifier. The feature extraction portion consists of five independent 3D convolution blocks. In each of the first two blocks, one stack of 3D convolution, 3D batch normalization (BN), 3D squeeze-and-excitation (SE) layers, and ReLU activation is followed by a 3D max-pooling operation. Each of the last 3 blocks contains two repetitions of 3D convolution, 3D batch normalization, 3D squeeze-and-excitation operation, and a ReLU activation. Block 3 and 4 ends with a 3D max-pooling operation but the 3D max-pooling operation of block 5 is abandoned to preserve enough size of the feature maps to generate the class activation map. The classifier is comprised of three dense layers, with the first two layers preceded by a dropout layer. The first dense layer is followed by the ReLU activation function whereas the second dense layer is followed by the sigmoid activation function, and the final dense layer is followed by the softmax function.

## 4 Discussion

This study investigated the performance of 3D VGG-based models on the classification of schizophrenia patients using structural MRI scans. The proposed model (SE-VGG11-BN) showcased superior performance and generality compared to the benchmark model in terms of sensitivity, specificity, accuracy, and AUC in both of our experiments. Furthermore, the proposed model, SE-VGG11-BN, was interpreted with gradient class activation maps to visualize the brain regions critical for classification. The important regions for classification involved subcortical and ventricular areas; these were in line with the findings in the previous literature.

### 4.1 The Superior Performance of the Proposed Model Against the Benchmark Model

SE-VGG-11BN exhibited better performance than the benchmark model. Several factors may have contributed to this result. Firstly, in contrast to the benchmark model, the proposed model contains squeeze-and-excitation (SE) blocks, which can capture important patterns across all channels after each convolutional operation. Secondly, the input of the proposed model was only down-sampled by a factor of two as opposed to the benchmark model, which used a larger factor of eight. Severely down-sampling the data likely negatively impacted model performance as lower-resolution inputs may have lost important information relevant to schizophrenia classification. Thirdly, we applied skull-stripping as part of our data preprocessing pipeline, given that the skull holds limited clinical correspondence to schizophrenia. The benchmark model used T1W WH scans, which may have unnecessarily confused the model with irrelevant features from the skull.

### 4.2 Interpretation of the Proposed Model's Grad-CAM

The class activation map of the proposed model (SE-VGG-11BN) reveals that the subcortical regions and ventricular areas provide the most discriminative structural information for schizophrenia classification. This result is consistent with two recent meta-analyses considering changes in regional brain structure volume [2] and shape [1] associated with schizophrenia. The first study examining the heterogeneity and homogeneity of regional brain structure in schizophrenia found that mean volumes were significantly reduced for the temporal lobe, frontal lobe, anterior cingulate cortex, and the subcortical regions including the thalamus, hippocampus, and amygdala; whereas, mean volumes of the lateral and third ventricles were significantly increased in patients [2]. The second meta-analysis investigating changes in subcortical brain shape associated with schizophrenia studied T1-weighted structural MRI scans from 2,833 individuals with schizophrenia and 3,929 healthy control participants contributed by 21 worldwide research groups participating in the ENIGMA Schizophrenia Working Group [1]. This study revealed more-concave-than-convex shape differences in the hippocampus, amygdala, accumbens, and thalamus in individuals with schizophrenia compared with control participants, more-convex-than-concave shape differences in the putamen and pallidum, and both concave and convex shape differences in the caudate. Patterns of exaggerated asymmetry were observed across the hippocampus, amygdala, and thalamus in individuals with schizophrenia compared to control participants, while diminished asymmetry encompassed the ventral striatum and ventral and dorsal thalamus. Notably, the hippocampus, a region found to be remarkably related to schizophrenia progression [32-36], is also included in the activation regions. Findings from our deep learning-based study suggest that common mechanisms may contribute to volume and shape variability across multiple subcortical regions and ventricular areas, which may enhance our understanding of the nature of network disorganization in schizophrenia.

### 4.3 Limitations and Future Work

There are certain limitations associated with the application of 3D CNN to schizophrenia classification. Firstly, the high computational cost during training caused by data with high dimensionality and large numbers of trainable parameters in the model may constrain the development of 3D CNN models. In this study, we down-sampled the raw data to reduce the GPU memory workload and preserved as many details as possible, simultaneously using a small down-sampling factor of two. The complexity of the models is also largely limited by the GPU memory requirement. Secondly, the sample size used in this study is relatively modest, especially for the 3D CNN network training. This most likely results in less efficient feature extraction and lower model generalizability. Introducing other high-quality labeled datasets coupled with data augmentation and effective image synthesis of new data could help model feature extraction and generalization via the introduction of more inter-subject anatomical variability and data quality deviation across different sites. Thirdly, registration error and down-sampling may ignore certain subtle anatomical differences and low-level contextual features potentially relevant to schizophrenia classification. Lastly, the training strategy of the model could be further improved. For instance, the proposed CNN model is trained from scratch, but applying and fine-tuning a pre-trained model on our data could further improve model performance. There is evidence suggesting that this approach may enhance performance by reducing the cost of a more computationally complex training stage [37].

A desirable future application of deep learning includes addressing the clinically more pressing question of discriminating schizophrenia from other psychiatric disorders, such as major depression and bipolar disorder. The partial overlap of genetic and similar symptoms characteristic between schizophrenia and other major psychiatric disorders make this task very challenging, even for clinicians. In fact, meta-analyses of transcriptomic studies covering five major psychiatric disorders found an overlap in polygenic traits and global gene expression patterns [38]. Moreover, the symptoms of schizophrenia also overlap with other psychiatric disorders such as major depressive disorder, schizoaffective disorder, and post-traumatic disorder [39, 40]. Though our proposed approach could differentiate schizophrenia patients from healthy controls using T1W structural MRI data, there is currently no objective method able to classify schizophrenia from other similar neuropsychiatric disorders. Further research in this area may elucidate the mechanisms and features underlying the brain structural alterations in different psychiatric disorders.

## 5  Data and Code Availability Statement

The T1W MRI scans used in this project are available from the SchizConnect database, http://www.schizconnect.org. The code used in this project is proprietary. The preprocessing script and the deep learning model are available upon request to the corresponding author. The code for this project is © 2022 The Trustees of Columbia University in the City of New York. This work may be reproduced and distributed for academic non-commercial purposes only.